\documentclass[11pt,twoside]{article}
\usepackage{./asp2014}
\aspSuppressVolSlug \resetcounters

\begin{document}

\title{AR Scorpii is a new white dwarf in the ejector state}
\author{N.G.\,Beskrovnaya$^{1}$ and N.R.\,Ikhsanov$^{1,2,3}$}
\affil{$^1$Pulkovo Observatory, Saint-Petersburg,  Russia;
\email{beskrovnaya@yahoo.com}} \affil{$^2$SAO RAS, Nizhni Arkhyz, Karachai
Cherkessia, Russia} \affil{$^3$Saint-Petersburg State University,
St.\,Petersburg,   Russia}

\paperauthor{Nina~Beskrovnaya}{beskrovnaya@yahoo.com}{ORCID_Or_Blank}{Pulkovo
Observatory}{Author1 Sector of Stellar Evolution}{Saint
Petersburg}{State/Province}{196140}{Russia}
\begin{abstract}
Marsh et~al. (2016) have recently reported the discovery of a radio-pulsing
white dwarf in the cataclysmic variable AR\,Sco. The period of pulsations
which are also seen in the optical and UV is about 117 seconds. High intensity
of pulsing radiation and non-thermal character of its spectrum leave little
room for doubt that the white dwarf in AR\,Sco operates as a spin-powered
pulsar and, therefore, is in the ejector state. We show that this system is
very much resembling a well-known object AE\,Aqr. In both systems the compact
components are spin-powered and have relatively strong surface magnetic field
of order of 100-500 MG. They originated due to accretion spin-up in the
previous epoch during which the magnetic field of the white dwarf had
substantially evolved being initially buried by the accreted matter and
recovered to its initial value after the spin-up phase had ended.
\end{abstract}
\section{Introduction}
Information about unique properties of AR\,Sco has been published quite
recently \citep{Marsh2016} and immediately excited much interest. This is a
close binary with the orbital period of $\sim 3.56$\,h containing a red dwarf
of spectral type M and a white dwarf spinning at the period of $\sim 117$\,s.
The most peculiar feature of this object is that the spin period of the white
dwarf is increasing so rapidly that its spin-down power exceeds the bolometric
luminosity of the system emitting electromagnetic radiation from radio to
X-rays. Moreover, the white dwarf is a radio-pulsar as well as a pulsar in the
IR, optical and UV spectral domains. The intensity of pulsed emission in the
optical region exceeds the total emission of the stellar components. The
spectrum of radio-emission is non-thermal and characterized by anomalously
high brightness temperature. \citet{Marsh2016} have concluded that the white
dwarf in AR\,Sco is  spin-powered and in the ejector state.

In spite of all unique properties, AR\,Sco is not the only system containing a
white dwarf in the ejector state, i.e. spending its  rotation energy according
to the canonical model of a radio-pulsar at a rate which exceeds the observed
source luminosity. It shares this property with the white dwarf of AE\,Aquarii
(AE\,Aqr) which is spinning at the period of 33\,s and exhibits a period
derivative of $5.64 \times 10^{-14}\,{\rm s}{\rm s}^{-1}$
\citep{de-Jager-etal-1994}.   Its spin-down power under these conditions
exceeds the bolometric luminosity of the system (observed in the wide range
from radio to X-rays) and can be explained within the conventional
spin-powered pulsar model provided the magnetic moment of the white dwarf is
$\mu \sim 10^{34}\,{\rm G}{\rm cm}^3$, which implies the surface magnetic
field of $50-100$\,MG \citep{Ikhsanov-1998, Ikhsanov-Biermann-2006}.

\section{Magnetic field of the white dwarf  in AR\,Sco}

In the case of AR\,Sco the spin-down power can be inferred from observations
as
\begin{equation}
L_{\rm sd} = I \omega_{\rm s} \dot{\omega}_{\rm s} \simeq 9.7 \times
10^{32}\ {\rm erg\,s^{-1}} \, \times \, I_{50}\ P_{117}^{-3}\
\left(\frac{\dot{P}}{3.9 \times 10^{-13}\,{\rm s\,s^{-1}}}\right),
  \end{equation}
where $I_{50}$ and $P_{117}$  are the  moment of inertia and the
spin period of the white dwarf in units of
$10^{50}\,\text{g}\,\text{cm}^2$ and 117\,s, respectively. On the
other hand, the spin-down power of a spin-powered pulsar with the
angular velocity $\omega$ and dipole magnetic moment $\mu$ can be
expressed following \citet{Spitkovsky-2006} as $L_{\rm rot} \approx
(\mu^{2}\omega^{4}/c^{3})(1+\sin^{2}\alpha)$, where $\alpha$  is the
angle between the magnetic and rotation axes. Solving equation
$L_{\rm sd} = L_{\rm rot}$ for $\mu$ and assuming $\alpha \sim
\pi/2$ one finds a lower limit to the dipole magnetic moment of the
white dwarf in AR\,Sco within the spin-powered white dwarf scenario
as
 \begin{equation}
 \mu \sim 5.6 \times10^{34}\, {\rm G\,cm^3} \,\times \,
 I_{50}^{1/2} \, P_{117}^{1/2} \,
 \left(\frac{\dot{P}}{3.9\times 10^{-13}\,{\rm s\,s^{-1}}}\right).
 \end{equation}
 This implies that the magnetic field strength at the equator of the white dwarf of
 mass $M_{\rm wd}$ and radius $R_{\rm wd}$ is
\begin{equation}
B\sim 150\,{\rm MG}\,\times k_{0.33}\,\left(\frac{M_{\rm wd}}{0.8
M_{\sun}}\right)^{1/2}\,\left(\frac{R_{\rm wd}}{7\times 10^{8}\,{\rm
cm}}\right)^{-2},
\end{equation}
where the moment of inertia of the white dwarf $I=k^{2}R_{\rm wd}^{2}M_{\rm
wd}$ and the radius of gyration $k_{0.33}=k/0.33$ is normalized following
\citet{Frank-etal-2002}.

\section{Origin  of AR\,Sco}
\vspace{-1mm}
\subsection{Spin-up epoch}
A possible formation of the ejector white dwarf in the process of
binary evolution was considered in detail in a previous paper
\citep{Ikhsanov-Beskrovnaya-2012} dedicated to the overview of
peculiar properties and origin of AE\,Aqr. The developed approach
can be used here since the white dwarf in AR\,Sco is currently
spinning down at the time-scale $t_{\rm sd} \simeq P_{\rm
s}/2\dot{P} \simeq 10^7$\,yr, while its age determined by the
cooling time is much longer. Indeed,  the cooling time of the white
dwarf with the surface temperature  $T_{\rm wd} \sim 10\,000$\,K and
mass $M_{\rm wd} \sim (0.8-1.3)\,M_{\sun}$  is $\ga 10^9$\,yr
\citep{Schoenberner-etal-2000}. A natural conclusion is  that fast
rotation of the white dwarf is not caused by the singularity of its
birth but is a consequence of peculiar binary evolution
incorporating the stage of rapid spin-up of the white dwarf due to
high-rate accretion onto its surface.

A white dwarf can undergo a significant  accretion-driven spin-up only under
condition $\dot{M}_{\rm su} > \dot{M}_{\rm cr}$, where $\dot{M}_{\rm cr}
\simeq 10^{-7}\,M_{\sun}/$yr is a critical  rate which provides a stable
burning of hydrogen in the matter accreted onto its surface \cite[see,
e.g.][and references therein]{Nomoto-etal-2007}. Otherwise, the situation
would resemble the dwarf novae in which the angular momentum is effectively
carried away from the white dwarf by the expanding envelope formed in the
novae outburst due to thermonuclear runaways on its surface
\citep{Livio-Pringle-1998}. For $\dot{M}_{\rm su} \sim \dot{M}_{\rm cr}$
  the duration of the spin-up stage is \citep{Ikhsanov-1999}
\begin{equation}\label{dur}
  t_{\rm su} \geq \frac{2\pi I}{\dot{M}_{\rm su}\ \sqrt{GM_{\rm wd} R_{\rm
m}}}\ \left(\frac{1}{P} - \frac{1}{P_{\rm 0}}\right) \simeq 1.2
\times 10^5\,{\rm yr}\, \times\, I_{50} \dot{M}_{19}^{-1}
M_{0.8}^{-1/2} R_9^{-1/2} P_{117}^{-1},
\end{equation}
where $P_{\rm 0} \gg P$ is an initial spin period of the white dwarf. Here
 $R_9$ is its magnetospheric radius, $R_{\rm m}$, at the end of the spin-up
epoch expressed in $10^9$\,cm, $M_{0.8}=M_{\rm wd}/0.8M_{\sun}$ and
$\dot{M}_{19} = \dot{M}_{\rm su}/10^{19}\,\text{g\,s}^{-1}$.
\vspace{-2mm}
\subsection{Screening of magnetic field}
%
In the process of accretion-driven spin-up the white dwarf could reach the
period $P\le 117\,{\rm s}$ only if its magnetospheric radius at the final
stage of the spin-up epoch, $R_{\rm su}$, did not exceed the corotation
radius, i.e. $R_{\rm su} \le R_{\rm cor}$, from which we estimate the magnetic
field strength of the white dwarf at the end of the spin-up stage as $B_{\rm
s} \leq B_0$, where
 \begin{equation}
 B_0 \simeq 10\,\text{MG} \, \times\,   M_{0.8}^{5/6} P_{117}^{7/6} \dot{M}_{19}^{1/2}
 \left(\frac{R_{\rm wd}}{7\times 10^{8}\,{\rm cm}}\right)^{-3}.
 \end{equation}
This means that the magnetic field strength of the white dwarf should
significantly vary in the course of AR\,Sco evolution. It decreased by a
factor of $\sim 50$ during the spin-up epoch and returned to its initial value
upon its completion. Such evolution of the magnetic field could result from
its screening by the accreting material \citep{Bisnovatyi-Kogan-Komberg-1974}.
A possibility to bury the magnetic field of accreting compact objects was
investigated for both the neutron stars \citep{Konar-Choudhuri-2004,
Lovelace-etal-2005} and the white dwarfs \citep{Cumming-2002}. Studies have
shown that under favorable conditions the surface magnetic field of a star can
decrease by a factor of 100 with its subsequent reemerging  due to diffusion
through the layer of accreted plasma.
\vspace{-2mm}
\subsection{Spin evolution of AR\,Sco: general scheme and conditions}
Following this approach we can outline the following scheme of the
AR\,Sco evolution. Prior to the stage of active mass transfer the
magnetic field strength on the surface of the white dwarf in AR\,Sco
was close to its current value and the mass exchange rate was
$\dot{M} \ll \dot{M}_{\rm su}$. As the normal component overfilled
its Roche lobe,  the mass exchange rate increased up to
$\dot{M}_{\rm su} \sim 10^{19}\,\text{g\,s}^{-1}$. This led to
decrease of the magnetospheric radius of the white dwarf down to
$R_{\rm m}^{\rm (0)} \leq   10^{10}\times
\mu_{34}^{4/7}\,\dot{M}_{19}^{-2/7} M_{0.8}^{-1/7}$\,cm (here
$\mu_{34}=\mu/10^{34}\,{\rm G\,cm^{-3}}$) and start of disk
accretion onto its surface provided $R_{\rm m}^{\rm (0)} < R_{\rm
cor}$. The latter condition could be satisfied if the initial spin
period of the white dwarf was $P_{\rm 0} \geq 10$\,min.

Due to the surface field of the white dwarf   screening by plasma accumulating
on its polar caps  the magnetospheric radius of the white dwarf was further
decreasing.   At the final stages of spin-up epoch the magnetic field strength
on the white dwarf surface did not exceed 10\,MG, that allowed spin-up of the
white dwarf to the current value of its rotation period.

Assuming that the magnetospheric radius of the white dwarf is
decreasing at the same rate as its corotation radius, one can
estimate the minimum possible spin-up time of the white dwarf with
account for screening of its magnetic field in the process of
accretion from equation $I \dot{\omega}_{\rm s} = \dot{M}_{\rm su}
\left(GM_{\rm wd} R_{\rm cor}\right)^{1/2}$ as
  \begin{equation}\label{delta_t_min}
\Delta t_{\rm min} =   \frac{(2 \pi)^{4/3} I}{\dot{M}_{\rm su} (GM_{\rm
wd})^{2/3} P_{\rm s}^{4/3}}.
   \end{equation}
%
%
%
After the stage of active mass exchange is finished the magnetic field of the
white dwarf is regenerating in the process of diffusion   through the layer of
degenerate matter. The timescale of this process can be evaluated  following
\citet{Cumming-2002} as
 \begin{equation}\label{tau_diff}
\tau_{\rm diff} \simeq 3 \times 10^8\,{\rm yr} \, \times \,
\left(\Delta M_{\rm a}/0.1\,M_{\sun}\right)^{7/5}.
 \end{equation}
For the magnetic field of the white dwarf to reemerge before its spin period
can significantly increase we should require   $\tau_{\rm diff} \leq t_{\rm
sd}$. Substituting expression (\ref{tau_diff}) and solving this inequality for
$\Delta M_{\rm a}$ we find for the parameters of AR\,Sco:
 \begin{equation}
\Delta M_{\rm a} \leq 0.009 \, M_{\sun} \, \times \, P_{117}^{5/7}
\left(\frac{\dot{P}}{3.9 \times
10^{-13}\,\text{s\,s}^{-1}}\right)^{-5/7}.
 \end{equation}
Taking into account that   the amount of matter accumulated on the white dwarf
surface during the spin-up stage can be estimated as $\Delta M_{\rm a} \ge
\dot{M}_{\rm su} \Delta t_{\rm min}$ and substituting (\ref{delta_t_min}), we
can get upper limit for the moment of inertia of   the   white dwarf in
AR\,Sco providing its evolution to be described in terms of accretion-driven
spin-up:
 \begin{equation}
 I \leq 2.1 \times 10^{50}\, {\rm g\,cm^2} \, \times \, P_{117}^{4/3}
 M_{0.8}^{2/3}.
 \end{equation}
This condition is satisfied \citep{Andronov-Yavorskij-1990} for any white
dwarf of the mass in the range $0.81-1.29\,M_{\sun}$ determined by
\citet{Marsh2016}.
%
This indicates that the origin of AR\,Sco can be explained within
the scenario in which the accretion-driven spin-up is accompanied by
screening of the magnetic field of the accretor initially proposed
for AE\,Aqr \citep[for futher discussion see
][]{Ikhsanov-Beskrovnaya-2012}.

\section{Acknowledgements}
NGB acknowledges the support of the RAS Presidium Program P-7. NRI
acknowledges the support of the Russian Science Foundation (grant
Nr.\,14-50-00043).

\end{document}